# Automated Brain Metastases Detection Framework for T1-Weighted Contrast-Enhanced 3D MRI

Engin Dikici, John L. Ryu, Mutlu Demirer, Matthew Bigelow, Richard D. White, Wayne Slone, Barbaros Selnur Erdal, and Luciano M. Prevedello

*Abstract*— **Brain Metastases (BM) complicate 20-40% of cancer cases. BM lesions can present as punctate (1 mm) foci, requiring high-precision Magnetic Resonance Imaging (MRI) in order to prevent inadequate or delayed BM treatment. However, BM lesion detection remains challenging partly due to their structural similarities to normal structures (e.g., vasculature). We propose a BM-detection framework using a single-sequence gadolinium-enhanced T1-weighted 3D MRI dataset. The framework focuses on detection of smaller (< 15 mm) BM lesions and consists of: (1) candidate-selection stage, using Laplacian of Gaussian approach for highlighting parts of a MRI volume holding higher BM occurrence probabilities, and (2) detection stage that iteratively processes cropped region-of-interest volumes centered by candidates using a custom-built 3D convolutional neural network ("CropNet"). Data is augmented extensively during training via a pipeline consisting of random gamma correction and elastic deformation stages; the framework thereby maintains its invariance for a plausible range of BM shape and intensity representations. This approach is tested using five-fold cross-validation on 217 datasets from 158 patients, with training and testing groups randomized per patient to eliminate learning bias. The BM database included lesions with a mean diameter of ~5.4 mm and a mean volume of ~160 mm³. For 90% BM-detection sensitivity, the framework produced on average 9.12 false-positive BM detections per patient (standard deviation of 3.49); for 85% sensitivity, the average number of false-positives declined to 5.85. Comparison analysis showed that the framework produces comparable BM-detection accuracy with the state-of-art approaches validated for significantly larger lesions.**

## I. INTRODUCTION

BRAIN metastases (BM) are disseminated cancer formations commonly originating from breast cancer, lung cancer, or malignant melanoma [1]. Detection of BM is a tedious and time-consuming manual process for radiologists, with no allowance for reduced accuracy; missed detections potentially compromise the success of treatment planning for the patient. Accordingly, computer-aided detection approaches have been proposed to assist radiologists by automatically segmenting and/or detecting BM in contrast-enhanced Magnetic Resonance Imaging (MRI) sequences, which is the key modality for the detection, characterization, and monitoring of BM. Methods utilizing traditional image processing and machine learning techniques, such as template matching [2][3][4], 3D cross-correlation metrics [5], fuzzy logic [6], level sets [7], and selective enhancement filtering [8] are reported to produce promising results. In recent years, Convolutional Neural Network (CNN) based approaches have started to be used extensively in a variety of medical imaging problems [9], and this holds great promise for BM evaluation.

To our knowledge, the application of a Deep Neural Network (DNN) to segmentation of BM in MRI datasets was first introduced by Losch et al. [10]. Besides analyzing the impact of different network structures on the segmentation accuracy, their study also showed that a DNN can produce comparable or even better results with respect to previously reported state-of-art approaches. However, a limitation of their approach was a significant reduction in accuracy for the segmentation of tumors with sizes below 40 mm³.

Charron et al. [11] used DeepMedic neural network [12] for segmenting and detecting BM in multi-sequence MRI datasets as input, including post-contrast T1-weighted 3D, T2-weighted 2D fluid-attenuated inversion recovery, and T1-weighted 2D sequences. The study involved investigation of the impacts of epoch, segment, and/or batch sizes on overall accuracy, thus providing a well-documented hyper-parameter optimization process. The BM considered in their study had a mean volume of 2400 mm³, and the system detected 93% of lesions whereas producing 7.8 average false-positive detections per patient.

Liu et al. proposed a modified DeepMedic structure, "En-

DeepMedic" [13], with the expectation of improved BM segmentation accuracy and higher computational efficiency. The approach was validated with both the BRATS database [14] and their post-contrast T1-weighted MRI collection of brain metastases with a mean tumor volume of 672 mm$^3$. The system yielded an average Dice similarity coefficient of 0.67, where the detection false-positive rate in connection to the sensitivity percentage is not reported.

More recently, Grøvik et al. [15] demonstrated the usage of 2.5D fully CNN, based on GoogLeNet architecture [16], for detection and segmentation of BM. Their solution utilized multiple sequences of MRI for each patient: T1-weighted 3D fast spin-echo (CUBE), post-contrast T1-weighted 3D axial IR-prepped FSPGR, and 3D CUBE fluid-attenuated inversion recovery. Their database included 156 patients, with testing performed on 51 patients. For the detection of BM, at 83% sensitivity, average number of false-positives per patient is reported as 8.3.

The motivation for our study is to provide a BM-detection framework for 3D T1-weighted contrast-enhanced MRI datasets that focuses on small lesions (≤15 mm) with an average volume of only ~160 mm$^3$. Such tiny lesions are difficult for even experienced neuroradiologists to detect, and missed lesions can lead to inadequate or delayed treatment. To our knowledge, no prior work focused on BM with volumes smaller than 500 mm$^3$. Detection of small lesions is particularly important given the clinical challenge they represent and due to recent paradigm shift in how these lesions are treated with radiation. In the past, patients with multiple intracranial metastases were treated with whole brain radiation, making detection of individual lesions not as crucial. However, due to long-term cognitive decline associated with whole brain radiation, recent radiation treatment regimens target individual lesion, consequently making detection of even a tiny lesion crucial for the appropriate treatment [17].

This report first provides the following components of the detection framework: (1) Candidate BM selection procedure, (2) training strategy, (3) data augmentation pipeline, and (4) CNN architecture. Next, a description of the medical data used in the study, BM data statistics, and its acquisition are presented. The evaluation criteria are then defined, and the motivation behind the use of the average number of false lesion-detections per patient in connection to the detection sensitivity is justified. Next, the results are provided based on a five-fold Cross-Validation (CV) executed on 217 datasets. Finally, it is concluded with: (1) Comparisons with other state-of-art techniques in the field, (2) summary of the novelties of the introduced study, (3) system limitations, and (4) future work indicators.

## II. METHODS AND MATERIALS

The BM-detection framework consists of two main components: (1) Candidate-selection step, and (2) a classification stage. First, the input MRI volume is processed using an information-theory based approach for detection of image points with high probability of representing BM. Next, volumetric regions centered by these candidate locations are iteratively fed into a custom-built CNN, *CropNet*, with extensive data augmentation, including rigid and non-rigid geometric transformations and intensity-based transformations. *CropNet* is a classification network, trained and validated to determine the probability of a given volumetric image to contain a BM. Algorithmic details of these stages are further described in the following subsections.

### A. Metastasis Candidate Selection

The visual appearance of metastatic masses can be generalized to blob-shaped formations either with relatively brighter or darker interiors (i.e., due to central necrosis). Blob-detection has been previously addressed using various generalized scale-space methods [18], including the Laplacian of Gaussian (LoG) approach [19]. In the proposed detection framework, LoG is utilized for detecting BM candidates for a given MRI volume as it: (1) Avoids image noise via its inherited Gaussian filtering properties, (2) holds few parameters to optimize, and (3) robustly detects BM candidates, with sensitivity reported in the Results section.

Yu et al. deployed LoG in the detection stage of their BM segmentation approach for MRI images [20], solidifying the applicability of LoG in the domain of our study. We further enhance the approach with sensitivity constraints and use it in candidate selection.

Given volumetric image data $V$, scale-space representation can be defined as,

$$L(x, y, z; s) = G(s) * V, \tag{1}$$

where $s$ is the scale, and $L$ gives the scale-space representation at $s$. Accordingly, the scale-normalized Laplacian operator is:

$$\nabla^2_{norm} L = s(L_{xx} + L_{yy} + L_{zz}). \tag{2}$$

Local optima of the above equation, which are maxima/minima of $\nabla^2_{norm} L$ with respect to both space and scale, represents the blob center positions [19].

The BM candidate-selection process aims to determine a set of image points that are closely located to the metastatic mass centers. Keeping the candidate list as short as possible is one of the main objectives for the process. However, the sensitivity of the framework needs to be maintained, which implies a comprehensive list of candidates. As these objectives are working against each other, the optimization process can be described as a minimax problem:

$$arg \ max_p(Sv(LoG(p, V), M)), \tag{3}$$
$$arg \ min_p(|LoG(p, V)|), \tag{4}$$

where $Sv$ defines the sensitivity of the system based on (1) $M$ representing the list of actual BM centers, and (2) $LoG(p, V)$ denoting candidate points selected for input volume $V$ with LoG parameters of $p$. As the sensitivity of the system is the major criterion in this study, we propose a solution where the sensitivity portion of the equation is constrained as

$$arg\ max_{p,S_{v \geq \theta}}\big(S(LoG(p,V),M)\big), \qquad (5)$$

with $\theta$ giving the minimal allowed sensitivity (e.g., 95 percent), and $p$ is found via grid search [21] constrained with Equation-4.

### B. Network Training

The DNN described in the following subsection aims to classify each BM candidate as positive (implies that the candidate point holds high probability for being a center of metastatic mass) or negative. The candidate-selection procedure using the LoG approach with sensitivity constraint, as described previously, produces candidates in magnitudes of thousands (please refer to Results section for actual numbers). Fortunately, only a few of the computed candidates are actual BM. Thus, the network training should factor in highly unbalanced class representations. The proposed detection framework addresses this using (1) random paired data selection strategy, and (2) on the fly data augmentation stage aiming to represent the covariance of tumor representations using a stochastic methodology.

During the training of the DNN, at each batch iteration, a pair of positive and negative samples are selected from each dataset randomly, producing a batch of $2N$ samples where N is the number of training cases. Next, the given batch is augmented on the fly [22], and the DNN is trained with the augmented batch (see Fig.1, row A). The term "epoch" is not used in this definition; as in the proposed framework, the samples are processed in a random pair basis, whereas epoch commonly refers to complete pass through all training data.

The augmentation process is the key for the introduced detection framework's learning invariance. The BM sample count is a small fraction of the total amount of samples—the learning process heavily depends on properly generalizing intensity and shape variations of BM. The importance of data augmentation for general computer vision and similar medical imaging scenarios are further described in [23] and [24], respectively. The detection framework deploys an augmentation pipeline consisting of random (1) elastic deformation, (2) gamma correction, (3) image flipping, and (4) rotation stages (see Fig.1, row B). In the following subsections, technical details for random elastic deformation and random gamma correction augmentations are provided. Next, the CNN, which processes the augmented positive and negative sample volumes, is further described.

#### 1) Random Elastic Deformations

Plausible non-rigid augmentations of the BM regions are produced by generalizing the random elastic deformation field generation algorithm proposed by Simard et al [25] from 2D to 3D. For a given volumetric image data $V$, random displacements fields $\Delta V_x$, $\Delta V_y$, and $\Delta V_z$ are defined, where each of these has similar dimensions as $V$, and their voxels hold random values picked from a uniform distribution defined in the range of $[0,1]$. Next, these random fields are smoothed with a zero-centered Gaussian kernel with a standard deviation of $\sigma$ (defined in mm). Finally, the deformation field is scaled with an elasticity coefficient $\alpha$. Choice of $\sigma$ causes elastic deformation to be (1) pure random with $\sigma \leq 0.01$, and (2) almost affine with $\sigma \geq 5$, whereas $\alpha$ determines the magnitude of the introduced local deformations (Fig. 2).

The usage of elastic deformations in the augmentation stage is crucial for the proposed framework, as it facilitates the generation of a conceivable BM shape domain. However, the algorithm needs to be used with well-tested parameters to ensure the viability of the augmented BM samples. In their paper, Simard et al. suggest the usage of $\sigma = 4$ and $\alpha = 34$, as it yielded the best results in their analyses. Our framework adopted those optimal parameters after visual inspections by a medical expert.

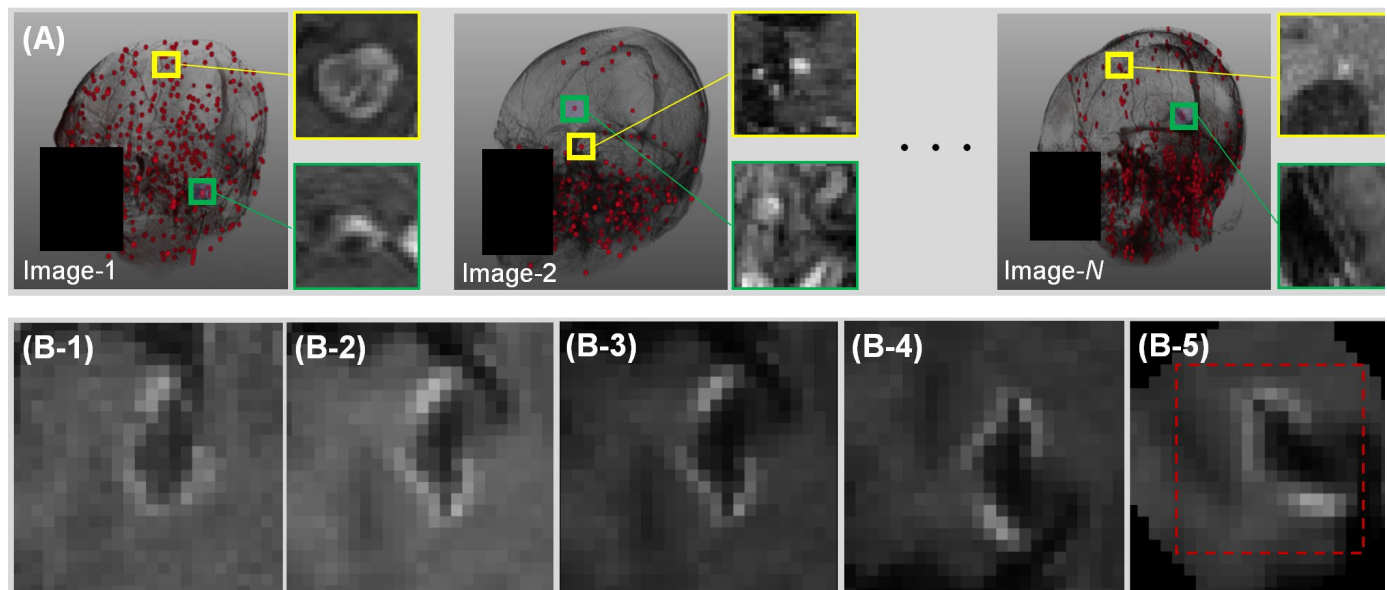

Fig. 1. Row A: Compilation of the positive & negative pair batch is represented. Positive and negative samples (shown with yellow and green rectangles respectively), are selected from BM candidates shown with red spheres in each dataset. Row B: Each positive sample goes through augmentation process: (B-1) mid-axial slice of an original cropped sample, (B-2) random elastic deformation is applied, (B-3) random gamma correction is applied, (B-4) sample volume is randomly flipped, and (B-5) sample volume is randomly rotated. The middle part of the randomly cropped volume, shown with a dashed red square in B-5, is used for the training. Face regions are covered to protect patient privacy.

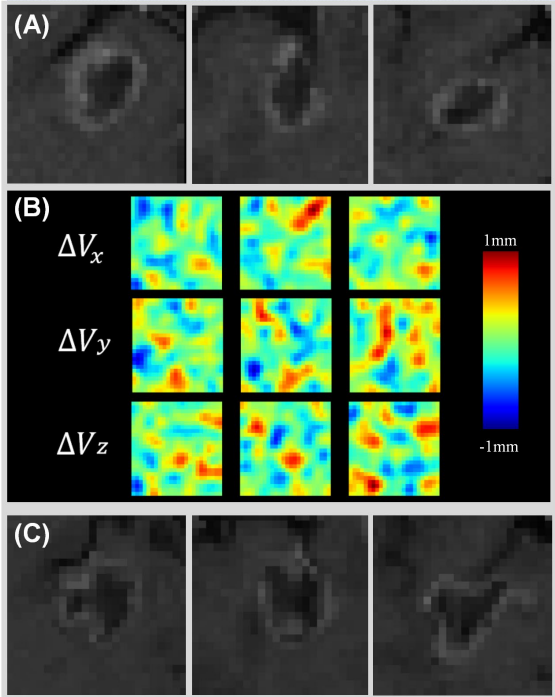

Fig. 2. The original cropped volume of a metastatic tumor mass (A), random displacement fields for x, y and z axes (B), and the corresponding deformed volume (C) are each shown from mid-axial, sagittal, and coronal views.

### 2) Random Gamma Corrections

In MRI, tissues do not have consistent intensity ranges, such as in computed tomography. Usage of bias field correction might improve the predictability of tissue intensities. However, its success is limited due to machine-dependent parameters [26]. Medical image processing algorithms, both information-theory and DNN based, benefit from understanding the probabilistic distributions of tissue intensity values. One way to achieve this goal is the normalization of image intensities in MRI to represent the target tissues with predefined intensity ranges [27]. Using even order derivatives of the histogram [28], Gaussian intensity normalization of selected tissues [29], and utilizing the median of intensity histogram [30] are some of the approaches introduced for that purpose. However, these methods are shown to be prone to errors as they aim to define approximations to non-linear intensity matching problems. The region-based approach [31], is shown to be effective, as it divides the spatial domain into smaller regions to address this limitation via piecewise linear approximations.

In the proposed framework, a novel form of the region-based strategy is introduced; random gamma corrections are applied to cropped volumetric regions during the augmentation stage [32]. Accordingly, the framework (1) does not make any assumptions about the histogram shape or intensity characteristics of given MRI datasets, and (2) avoids losing or corrupting potentially valuable intensity features, which is a common disadvantage of image intensity normalization-based methods.

Gamma correction of given volumetric data is given by,

$$V_G = V_N^{(1/\gamma)}, \tag{6}$$

where $V_N$ is the intensity scaled volumetric image data in $[0,1]$ range, $\gamma$ is the gamma value, and $V_G$ is the gamma-corrected volumetric image data, which is also intensity scaled (see Fig. 3).

In the detection framework, the gamma correction augmentations are utilized by randomly picking $\gamma$ values from a uniform distribution defined in $[0.8, 1.2]$ range, determined empirically by investigating the visual appearance of gamma-corrected volumetric regions.

### 3) Network Architecture

The CNN introduced in this study (i.e., *CropNet*) has an input layer with an isotropic-sampled volumetric region of interest (ROI), where each voxel represents 1 mm³. Please note that the input volume's edge length is used in model naming, such as *CropNet*-[*c*]mm, where *c* is the volume's edge length in mm. Besides, the model follows a typical contracting path structure: Each resolution level is formed using stacked blocks each consisting of convolution, rectified linear activation unit (ReLU) and dropout layers. Block count per resolution level is another configurable parameter for the introduced network, hence, included in the naming convention as *CropNet*-b[*B*], where *B* denotes the number of blocks per level. The network's downsampling is performed via $2 \times 2 \times 2$ max-pooling, followed with channel doubling. The output is a one-dimensional scalar produced via the sigmoid activation layer, which holds value in the range of $[0,1]$ representing the likelihood of a given ROI to contain a metastatic mass. The network's convolution layers are initialized using Glorot uniform initializer as described in [33].

In Fig.4, the formation of network architecture is illustrated for two blocks and 32 mm edge length (*CropNet*-b2-32mm), thus the reader can associate naming convention with the CNN. As described in the Results section, the study employs

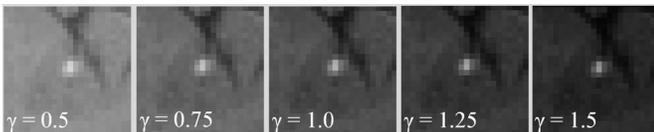

Fig. 3. The effects of gamma correction on region centering ~2.2mm diameter metastasis (mid axial slice of a cropped 3D volume). Please note that γ=1.0 represents the original image.

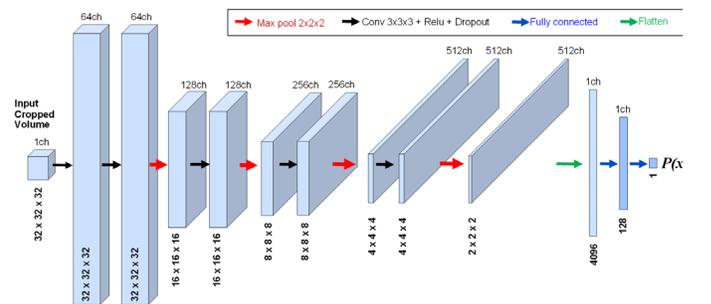

Fig. 4. *CropNet*-b2-32mm: Input of this CNN is 32mm x 32mm x 32mm isotropic region-of-interest, and each resolution level consists of two identical blocks, where the output is a scalar in range of [-1, 1].

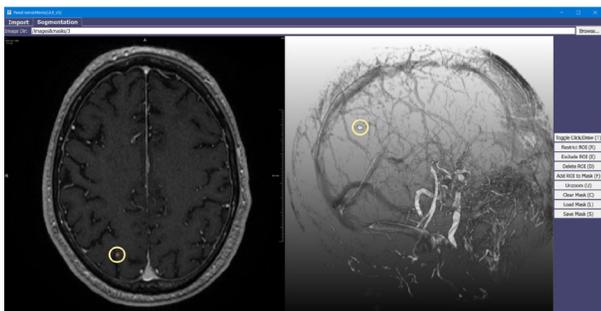

Fig. 5. The screenshot of the custom tool allowing medical experts to create, edit, save/load segmentation masks of BMs in MRI images. The tool provides 2D axial view, 3D view, and various manual editing tools.

16 mm version of CropNet, as (1) the target objects have diameters smaller than 15 mm, and (2) *CropNet*-b2-16mm produced comparable performance and allowed faster training with respect to its higher edge length versions (i.e. 32 and 64 mm).

## III. DATABASE

### A. Data Collection

This retrospective study was conducted under Institutional Review Board approval with waiver of informed consent (institutional IRB ID: 2016H0084). A total of 217 post-gadolinium T1-weighted 3D MRI exams were collected from 158 patients: 113 patients with a single dataset, 33 patients with 2 datasets (i.e. one follow-up examination), 10 patients with 3 datasets, and 2 patients with 4 datasets. Two of the major study selection parameters were that (1) none of the datasets involved lesions with diameter of 15 mm or larger, and (2) motion degraded studies were included.

Ground-truth BM segmentation masks were prepared by a radiologist, using a custom-built tool for the project [34]. The tool was developed using MeVisLab 2.8 (medical image processing and visualization framework developed by MeVis Medical Solutions AG), and it allows users to load volumetric MRI datasets, manually delineate the borders of BM, and edit the existing segmentation masks if needed (see Fig.5).

### B. Brain Metastases

The database included 932 BMs where, (1) mean number of BMs per patient is 4.29 ($\sigma = 5.52$), median number per patient is 2, (2) mean BM diameter is 5.45 mm ($\sigma = 2.67$ mm), median BM diameter is 4.57 mm, and (3) mean BM volume is 159.58 mm$^3$ ($\sigma = 275.53$ mm$^3$), median BM volume is 50.40 mm$^3$. Fig.6 (A, B and C), provides the histograms for each of these distributions.

For better understanding of the localization of BMs included in our database, all BMs are registered on a reference MRI image, and the probability density function is generated for multiple projections in Fig.6 (D). The volumetric registration for this illustration is performed by maximizing the mutual information between the reference MRI volume, and the rest of the volumes in the database iteratively, maximizing:

$$I(V_C, V_{Ref}) = H(V_{Ref}) - H(V_{Ref}|V_C,) \qquad (7)$$

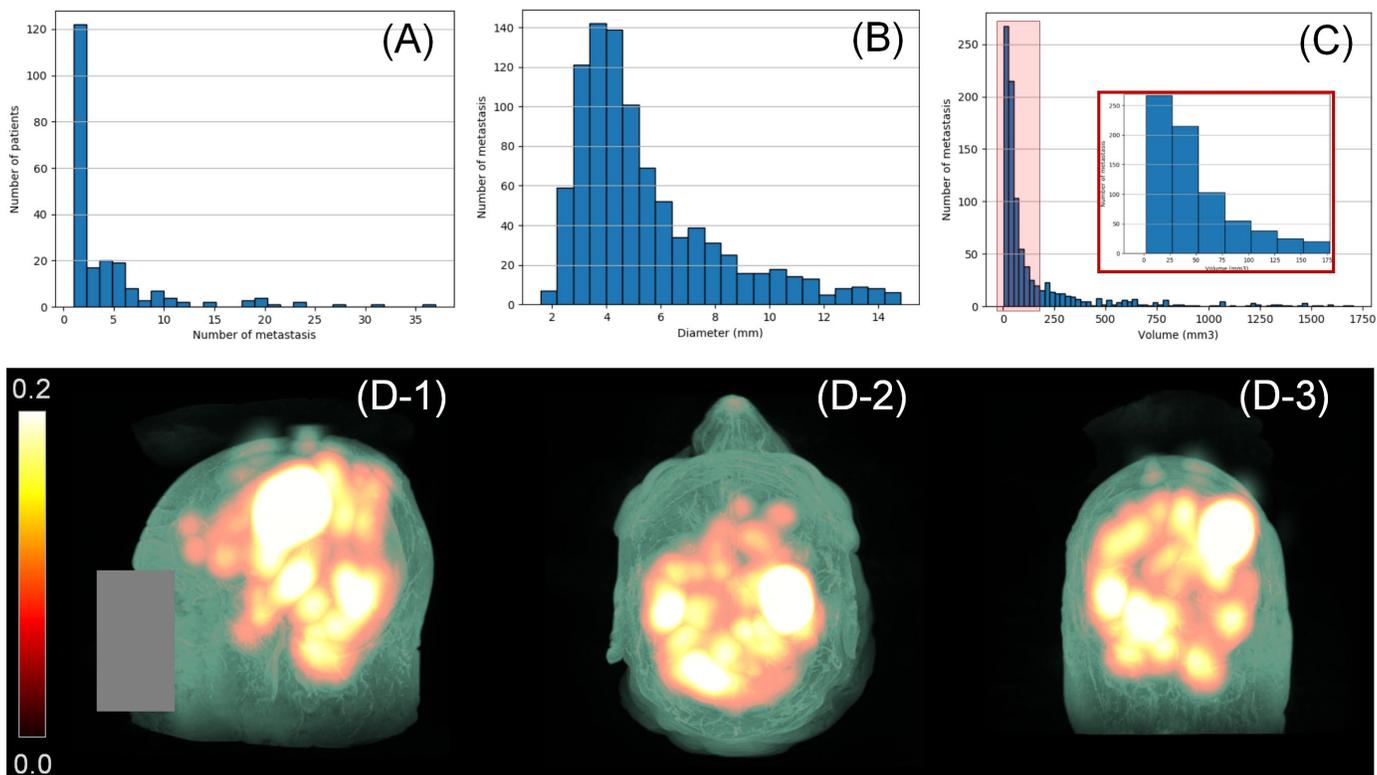

Fig. 6. The histograms for (A) number of BM per patient, (B) diameters of BM, and (C) volumes of lesions in BM database are shown. Below, the BM probability density function's projections on left sagittal (D-1), axial (D-2), and coronal (D-3) planes are provided. Face region is covered to protect patient privacy.

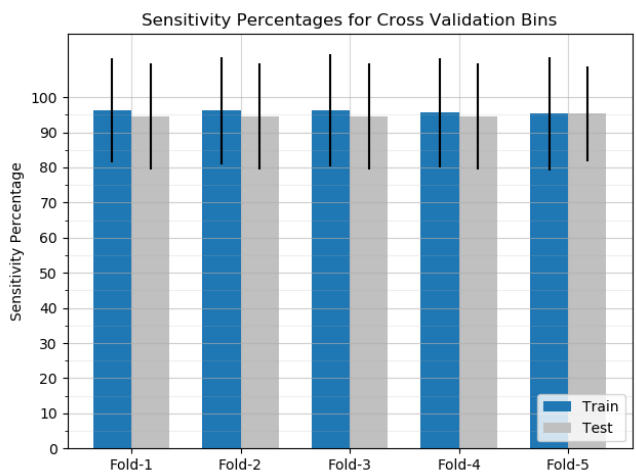

Fig. 7. Candidate selection procedure's sensitivity percentages for each fold's training (blue) and testing (silver) groups are represented. Sensitivity standard deviations are also shown with bold lines on each block.

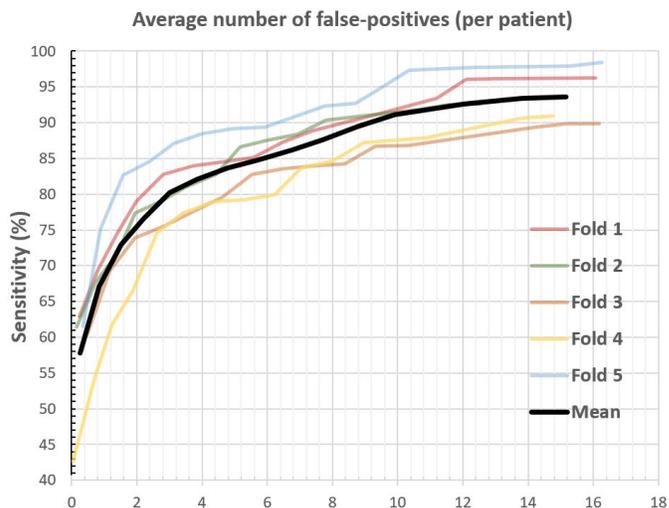

Fig. 8. Average number of false-positives per patient (i.e., wrongly detected BM lesions for each patient) in relation to the sensitivity is illustrated for each CV fold. The mean curve (shown with black) represents the average of the CV folds.

where $V_C$ is the floating volume (i.e. any volume picked from the database), $V_{Ref}$ is the reference volume, $H(V_{Ref})$ is the Shannon entropy of the reference volume, and $H(V_{Ref}|V_C)$ is the conditional entropy. Rigid registration, optimizing translation and rotation parameters, is utilized in our visualization. The interested reader may refer to [35] for further details on mutual information's usage in medical image registration.

### C. Evaluation Metric

The clinical applicability of a BM-detection algorithm was assessed by measuring (1) the sensitivity and (2) the average number of false lesion-detections for a given sensitivity.

As a screening tool, sensitivity of the system is expected to be high: In a typical deployment scenario of a detection algorithm, the appropriate operating point, maximizing the sensitivity whereas minimizing the average false lesion-detections per patient, needs to be adjusted by a medical expert. Therefore, we plot our performance metrics (i.e. sensitivity vs average number of false-positive detections per patient - AFP) at various output threshold settings (~0 – low likelihood and ~1 – high likelihood of metastasis). Accordingly, state-of-art approaches[10][11][15] follow a similar reporting methodology.

### IV. RESULTS

The detection framework is validated using 5-fold CV. Folds are generated based on patient, which ensures each patient is located either in a training or testing group for each fold (e.g. datasets from Patient-A are all located either in training or testing group for fold-n) for eliminating the learning bias. Accordingly, the bins included datasets from 31, 31, 32, 32 and 32 patients, respectively. For each CV fold, four bins are used for the training and validation, and a single bin is used for the testing.

For the candidate selection stage of the framework, Laplacian of Gaussian parameters are optimized from the training bins with the constraint of setting minimal sensitivity to 95% (see Equation-5). These parameters include (1) minimal and maximal standard deviations for the Gaussian kernel, and (2) the absolute lower bound for scale-space maxima, also referred to as LoG threshold in the literature [19]. The candidate-selection procedure achieved (1) a mean sensitivity of 95.8, where the sensitivity for training and testing groups of each fold are represented in Fig.7, and (2) produced 72623 candidates on average ($\sigma = 12518$) for each 3D dataset. Processing time for each dataset is ~30.6 seconds (using a 3.5 GHz Intel Core i7-5930K CPU.

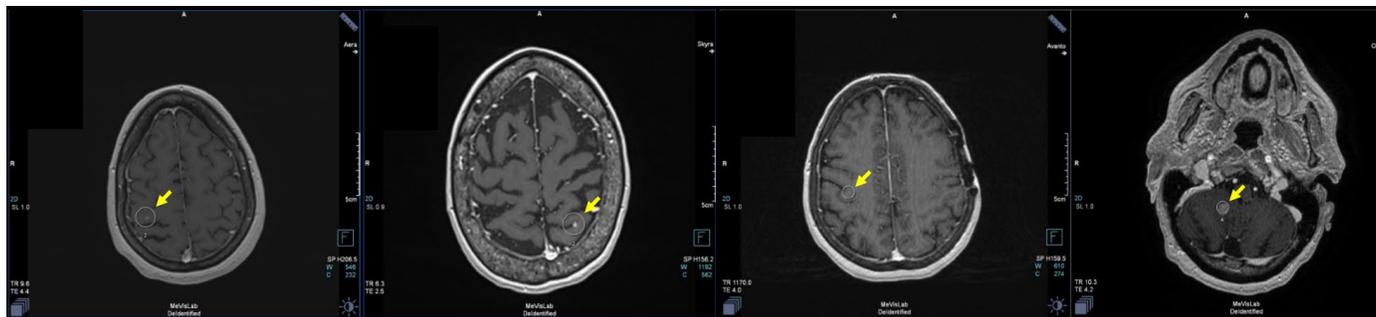

Fig. 9. The framework output; white circles centered by the BM detections are rendered (yellow arrows are added to the figure for the readers' convenience).

TABLE I
OVERVIEW OF BM DETECTION/SEGMENTATION STUDIES THAT USE CONVOLUTIONAL NEURAL NETWORKS

| Study | Patient # | Acquisition | BM diameter (mm) | BM volume (mm) | DNN Type | Validation Type | Sensitivity (%) | AFP |
|---|---|---|---|---|---|---|---|---|
| Losch et al. [10] | 490 | T1c MRI | NA | NA | Multi-scale ConvNet | Fixed train/test [e] | 82.8 | 7.7 |
| Liu et al. [13] | 490 | Multi seq.[a] | NA | Mean: 672 | En-DeepMedic | 5-fold CV | NA | NA |
| Charron et al. [11] | 182 | Multi seq.[b] | Mean: 8.1 Median: 7 | Mean: 2400 Median: 500 | DeepMedic | Fixed train/test [f] | 93 | 7.8 |
| Grøvik et al. [15] | 156 | Multi seq.[c] | NA | NA | GoogLeNet [d] | Fixed train/test [g] | 83 | 8.3 |
| **This study** | **158 [h]** | **T1c MRI** | **Mean: 5.4 Median: 4.6** | **Mean: 159.6 Median: 50.4** | ***CropNet*** | **5-fold CV** | **90** | **9.12** |

[a] 235 T1c MRI datasets, and 265 datasets from BRATS DB; including both T1c and T2-weighted Fluid-Attenuated Inversion Recovery (FLAIR) sequences.
[b] T1-weighted 3D MRI with Gd injection, T2-weighted 2D fluid attenuated inversion recovery MRI and T1-weighted 2D MRI sequences.
[c] Pre- and post-gadolinium T1-weighted 3D fast spin echo (CUBE), post-gadolinium T1-weighted 3D axial IR-prepped FSPGR (BRAVO), and 3D CUBE fluid attenuated inversion recovery (FLAIR) sequences.
[d] 2.5 dimensional fully connected convolutional net based on GoogLeNet.
[e] 440 training and 50 test cases.
[f] 164 training and 18 test cases.
[g] 100 training, 5 development and 51 test cases.
[h] 217 datasets are collected from 158 patients, CV folds are created patient-wise to ensure a patient can only exist either in training or testing group.

The framework contained *CropNet*-b2-16mm for processing the BM candidates and providing the final detection results. The network processed cubic ROIs with 16mm edges and each resolution level included two blocks with layers as described in Section.2. For each fold, *CropNet* is trained for 20000 batch iterations. The optimal version of the network is determined using the minima of moving validation loss average, computed over 30 batch iterations. On average, the training process took 11312 ($\sigma$ = 183) batch iterations to converge. The network's training time for each fold was ~3.5 hours using an NVIDIA 1080ti graphics card with 11 GB RAM.

The average number of false-positives (i.e. false lesion-detections) per patient (AFP) were computed in connection to the sensitivity of the framework for each CV fold, where the sensitivity of the framework was adjusted via setting a threshold at *CropNet*'s response. AFP was computed as 9.12 per patient with a standard deviation of 3.49 at 90 percent sensitivity. At lower sensitivity percentages, AFP was computed as 8.48 at 89%, 7.92 at 88%, 7.29 at 87%, 6.64 at 86%, and 5.85 at 85% (see Fig.8). Fig. 9 illustrates sample output screens for the deployed BM detection framework.

## V. DISCUSSION AND CONCLUSION

Table.1 provides an overview of the databases, acquisition types, neural network architectures, validation strategies and detection accuracies of some of the prominent CNN based BM-detection/segmentation approaches for 3D MRI, published over the recent years. From these, [11] and [15] requires multiple MRI sequences during the BM-segmentation/detection process, whereas [13] benefited from multiple sequences for the training and validation. Our framework trained and validated for a single type of MRI sequence, T1c, and requires only this type of study during its decision-making process.

The dimensional properties of the BM included in a detection study are critical for determining the clinical applicability of a proposed solution. This is due to the fact that smaller lesions are harder to identify even by highly trained neuroradiologists. Consequently, they may greatly benefit from a system trained and validated specifically for that type of data. As illustrated in Table.1, our study employed a BM database that included relatively smaller BM lesions compared with the referenced studies; the smallest BM average volume in comparable studies is 672 mm$^3$ [13], whereas the BM average volume in this study is only 159.58 mm$^3$.

BM-detection and segmentation databases used in our study and in other comparable studies (as shown in Table.1) are limited with respect to number of cases; they all consist of some hundreds of patients. Estimating the accuracies of such machine learning approaches, trained with a limited amount of datasets, can gain significantly from the usage of CV, as the method minimizes the error of algorithm's predictive performance evaluation [36]. Therefore, we found it valuable to emphasize the validation schemes of comparable studies in Table.1.

The study introduced multiple technical novelties: (1) Sensitivity constrained LoG BM-candidate selection, (2) random 3D Simard elastic deformation augmentations (Simard deformation field used for medical-image augmentation for the first time to our knowledge), (3) volumetric random gamma correction augmentations for MRI, and (4) a parametric CNN for processing cubic volumes of interests. More importantly, all of these components are put into a sound framework that can be utilized for various detection applications in medical imaging.

The limitations of the proposed solution are (1) its scope is currently limited to the detection; medical interventions requiring exact borders of detected tissues (such as stereotactic radiotherapy) may not benefit from the method, and (2) lesions with sizes larger than 15 mm would not be detected with a system trained with the parameters given in this paper. However, BM-candidate selection and *CropNet* parameters can be reset for the dimensional properties of the target tumors.

The performances of machine-learning algorithms, including the CNNs, heavily depend on their hyperparameter settings [37]. Accordingly, some of the BM-segmentation studies, such as [10] and [11], provided a set of analyses on parameter tuning. The introduced framework's performance also relies on proper setup of multiple parameters, including (1) edge length and the block count of *CropNet*, (2) random gamma correction range, and (3) elastic deformation parameters, which were found empirically and individually. Therefore, multivariate optimization of these in a future study might further improve the accuracy of the framework.

The introduced framework can be extended for segmentation of the metastatic mass lesions. The network's contracting layers can be appended with a symmetric set of expanding layers as in [24] or [22], and its loss function can be changed to Dice similarity coefficient, or another image segmentation metric [38], to perform segmentation. Alternatively, previously defined BM-segmentation algorithms can be modified to use the proposed detection framework in their preprocessing stages.

The proposed data augmentation pipeline uses random gamma transformations and elastic deformations to capture the BM intensity and shape variabilities. The strategy mimics the kernel density estimation with Parzen windows [39], as the probability densities of the BM with respect to intensity and shape are generated from a small set of actual BM (932 BM) and their ranged uniform variations to deploy an uniform kernel density. For density estimation problems, it is also common to use Gaussian kernel densities [39], which would translate to (1) using gamma corrections randomly picked from a normal distribution centered at 1 (i.e., $\gamma = 1$ gives the original image), and (2) elastic deformations randomly picked from a bivariate distribution centered at $(0,0)$ (i.e. $\sigma = 0$ and $\alpha = 0$ implies null Simard deformation field). The impact of kernel density function to the final accuracy is a topic for a future study.

This study introduced a novel BM detection framework that focused on small lesions. It is validated for its high sensitivity, and it produced relatively few false BM detections per patient. The results suggest that its detection performance is comparable with state-of-art approaches that are validated against significantly larger lesions. In addition to technical novelties introduced, the study focuses on an increasingly important field-of-research that is the detection of small BM in order to minimize challenges they pose to both radiologists and radiation oncologists in the identification and treatment of these lesions to ultimately improve patient care.